\documentclass[aps,onecolumn]{revtex4-2}
\usepackage[utf8]{inputenc}
\usepackage{amsmath}
\usepackage{hyperref}
\usepackage{mathdots}
\usepackage{dsfont}
\usepackage{amsfonts}
\usepackage{amssymb}
\usepackage{graphicx}
\usepackage{color}

\UseRawInputEncoding

\begin{document}

\title[Carleman-Grad approach to the quantum simulation of fluids]{Carleman-Grad approach to
the quantum simulation of fluids}
% Force line breaks with \\
\author{Claudio Sanavio}
\email{claudio.sanavio@iit.it}
\affiliation{Fondazione Istituto Italiano di Tecnologia\\
Center for Life Nano-Neuroscience at la Sapienza\\
Viale Regina Elena 291, 00161 Roma, Italy}

\author{Enea Mauri}
\affiliation{Fondazione Bruno Kessler\\
Via Sommarive 18, 38123, Povo (TN), Italy}

\affiliation{Fondazione Istituto Italiano di Tecnologia\\
Center for Life Nano-Neuroscience at la Sapienza\\
Viale Regina Elena 291, 00161 Roma, Italy}

\author{Sauro Succi}
\affiliation{Fondazione Istituto Italiano di Tecnologia\\
Center for Life Nano-Neuroscience at la Sapienza\\
Viale Regina Elena 291, 00161 Roma, Italy}

\date{\today}% It is always \today, today,

\begin{abstract}
We discuss the Carleman linearization approach to the quantum simulation of
classical fluids based on Grad's generalized hydrodynamics and compare it to
previous investigations based on lattice Boltzmann and Navier-Stokes formulations.
We show that the Carleman-Grad procedure exhibits intermediate properties
between the two. Namely, convergence of the Carleman iteration over a few tens of timesteps
and a potentially viable quantum circuit implementation using quantum linear algebra
solvers. However, both features still need substantial improvements 
to yield a viable quantum algorithm for fluid flows.    
\end{abstract}

\maketitle

\section{Introduction: quantum computing for fluids}

In recent years there has been a growing interest in exploring 
the potential of quantum computing for nonlinear problems in 
classical physics, and most notably the physics of 
fluids~\cite{steijl_quantum_2019,gaitan_finding_2020,steijl_quantum_2022,succi_quantum_2023,sanavio_quantum_2024}.

The potential of quantum computing for fluid flows is indeed mind-boggling
\cite{bharadwaj_quantum_2020,succi_ensemble_2024},
as readily appreciated by considering that the computational complexity 
of a turbulent flow at Reynolds number $Re$ scales like $Re^3$. 
Hence, using amplitude encoding, it requires a number of qubits
\begin{equation}
\label{QRE}
Q(Re) = 3 \log_2 Re \sim 10 \log_{10}Re.
\end{equation}

Current near-exascale electronic supercomputers can reach up to $Re \sim 10^7$ 
(a standard automobile), meaning $Q \sim 70$, which is well within the {\it nominal} capabilities
of current quantum hardware~\cite{ibm_roadmapsq}. 
Clearly, the optimistic relation described by Eq.~\eqref{QRE} needs to be 
scaled up by a factor of ten to a hundred, to acknowledge the gap between 
nominal (logical) qubits and functional (physical) ones, raising
the count to $Q \sim 10^3 \div 10^4$. While still beyond the 
capabilities of current quantum computers, this looks like a 
plausible target for the near to mid-term future. 

However, when it comes to fluids, in addition to handling decoherence and quantum noise, one 
has to devise suitable strategies to deal with two distinctive features with no counterpart
in (standard) quantum mechanics: nonlinearity and dissipation.
Many such strategies have been proposed over the last few years, but no
clearcut solution has emerged yet \cite{sanavio_quantum_2024}.
For instance, in \cite{gaitan_finding_2020} a quantum ODE solver is proposed, but no
specific implementation of the quantum oracle is presented. 
Another interesting possibility is to adapt variational quantum eigenvalue 
solvers to the fluid equations, to sidestep the difficult issue of time marching.
However, to date, it appears difficult to minimize the associated functional 
at the required level of accuracy~\cite{lubasch_variational_2020} for quantitative
fluid dynamic purposes.

Another promising approach is the one based on Carleman 
linearization~\cite{carleman_application_1932,sanavio_lattice_2024,sanavio_three_2024},
whereby the finite-dimensional nonlinear fluid problem is turned into an infinite-dimensional
linear one, to be truncated at a suitable level of the hierarchy. 

This is the method discussed in the present paper.      

\section{Carleman linearization of the logistic equation}

We begin by presenting the main ideas behind Carleman linearization by means 
of a simple pedagogical example in zero spatial dimensions, namely 
the following logistic equation:

% -----------------------------
\begin{figure}
    \centering
    \includegraphics[scale = 0.4]{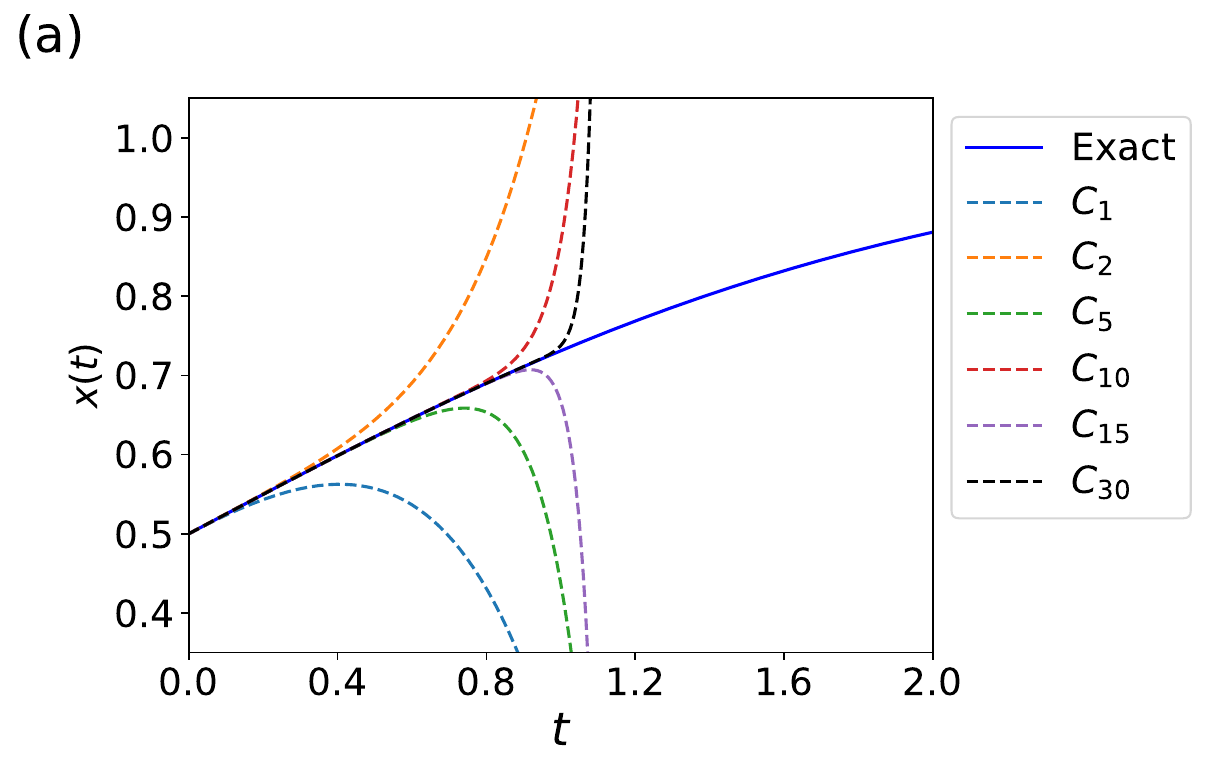}
    \includegraphics[scale = 0.4]{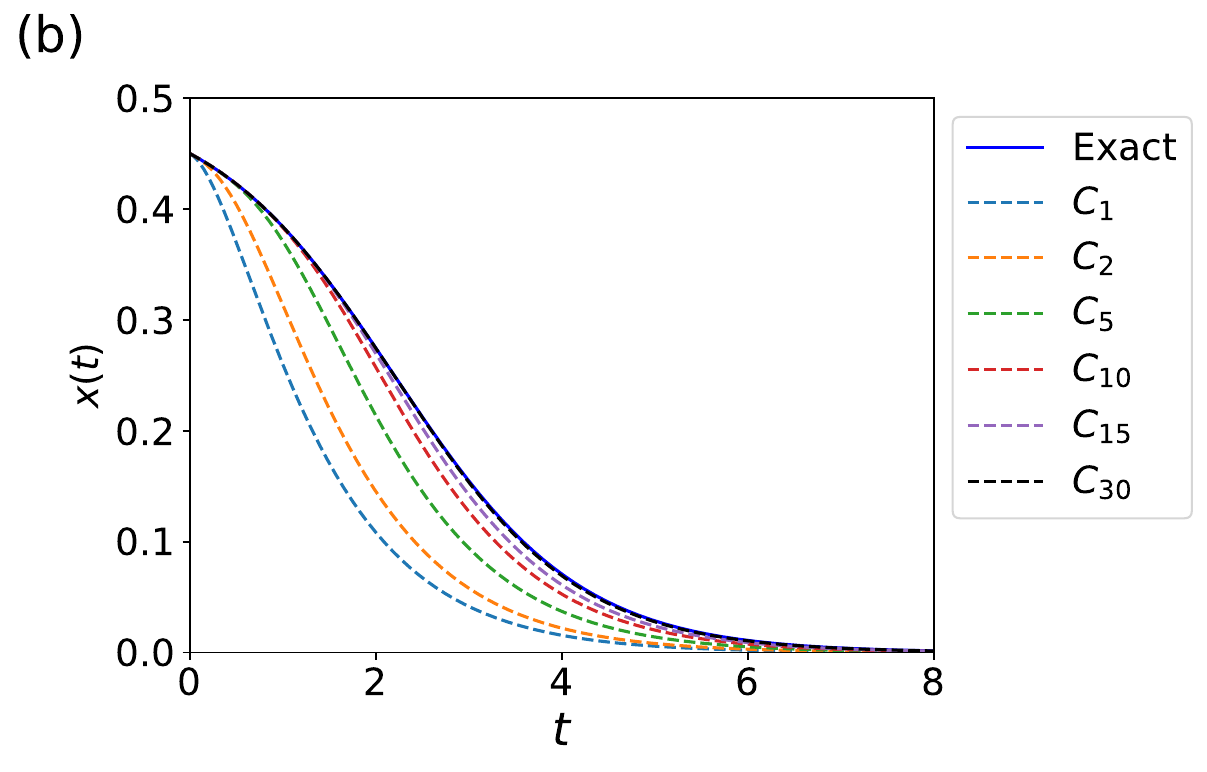}
    \caption{\label{fig:carleman_logistic} 
Convergence of the Carleman Linearization for the logistic equation at increasing
orders of the truncation. (a) The growing logistic equation, with $a=-1,b=-1$ and initial condition $x_0=0.5$. 
(b) The decaying logistic equation, with $a=1,b=1$ and initial condition $x_0=0.45$.} 
\end{figure}
% -----------------------------

\begin{equation}
\label{LOGI}
\dot x = - ax+bx^2,\qquad x(0)=x_0
\end{equation}
where $a>0$ is the decay rate and $b>0$ is the quadratic regeneration rate.

This equation admits a stable attractor $x_s= 0$ and an unstable 
one $x_u=1/R$, where $R=b/a$.
Hence, any initial condition such that $0<Rx_0<1$ is asymptotically
attracted to $0$, whereas for $Rx_0>1$, the solution blows up
in a finite time $t_{sing} = a^{-1} \log(\frac{Rx_0}{Rx_0-1})$, as it is readily
inferred by inspecting the exact solution:

\begin{equation}
\label{EXA}
x(t) = x_0 \frac{e^{-at}}{1-Rx_0 (1-e^{-at})},
\end{equation}

For the case $a<0$ and $b<0$, the attractors swap their role: 
$x_u=0$ becomes unstable and the stable one is $x_s=1/R$. 
Hence any initial condition $x_0>0$ is attracted to $1/R$, while any $x_0<0$
would diverge to minus infinity in a finite time.
It has to be noted that the decaying case, $a,b>0$ and the growing 
one $a,b<0$ turn one into another under the dual transformation $ x'=1/R-x$.
Hence they are basically equivalent.

The Carleman procedure consists in renaming $x^{(1)} \equiv x$ and 
$x^{(2)} \equiv x^2$, so that the logistic equation takes the linear form 
\begin{equation}
    \dot x^{(1)} = -a x^{(1)}+b x^{(2)} \text{ ,}\\
\end{equation}
with the equation for $x^{(2)}$ given by 
\begin{equation}
        \dot x^{(2)} = 2 x \dot x = -2a x^{(2)} + 2b x^{(3)} \text{ ,}
\end{equation}
where we defined $x^{(3)} \equiv x^3$.
Iterating the procedure to the $k$-th order delivers:
\begin{equation}
\dot x^{(k)} = -k (ax^{(k)}-bx^{(k+1)}).
\end{equation}

This is an endless hierarchy, which is then truncated at a given 
order $K$ by setting $x^{K+1}=0$, in the hope that the truncated solution 
can still capture the essential behaviour of the exact one over the desired timespan.

In practice, it can be shown that the Carleman hierarchy is just another way of
representing the exact solution as an infinite power series in the
term $Rx_0(1 - e^{-at})$. In fact, the truncation at order $K$ yields the  solution
\begin{equation}\label{eq:carleman_solution}
x(t)=x_0e^{-at}\sum_{k=0}^K\bigg{[}Rx_0(1-e^{-at})\bigg{]}^k.
\end{equation}
%For $Rx_0>1$, Carleman linearization fails to give an estimate of the solution 
%for $t>t_\text{lim}$, regardless of the truncation level, whereas
In the stable regime, $Rx_0<1$, higher truncation levels result in a better 
approximation, as shown in Fig.~\ref{fig:carleman_logistic}(b), with 
parameters set $a=1,b=2$ and $x_0=0.45$ . 

The growing case, albeit analytically equivalent under the duality 
transformation mentioned earlier on, shows a different behaviour.
It is straightforward to see that for the growing case $a,b<0$, the radius 
of convergence of the series in Eq.~\eqref{eq:carleman_solution} is given by
\begin{equation}
t_{\lim} = \frac{1}{|a|}\log\bigg{(}\frac{1+Rx_0}{Rx_0}\bigg{)}.
\end{equation}

\noindent This behaviour is shown in Fig.~\ref{fig:carleman_logistic}(a) for the case 
$a=-1,b=-1,x_0=\frac{1}{2}$, where each successive iteration expands the 
time horizon of convergence, until the 
time horizon $t_{\text{lim}}=\log(3) \sim 1.098 $ is reached.
As mentioned earlier on, the decaying and growing cases are dual, but the  
the Carleman procedure is not able to catch this equivalence, since
the growing case is restricted by the convergence 
radius of the series~\eqref{eq:carleman_solution}.
As a result, the Carleman expansion fails to capture the saturation regime.
 
Despite its simplicity, the logistic equation provides a few useful
hints for the Carleman linearization of the fluid equations.
First, like fluids, it presents a quadratic non-linearity. 
Second, it shows  that the time horizon of the Carleman series around an unstable attractor
depends on the strength of the nonlinearity via the dimensionless 
group $Rx_0$, namely the ratio of the initial to the time asymptotic condition $1/R$.   
Third, the logistic equation bears a close similarity
to the collision operator of kinetic equations.
This observation prompted the earliest attempts to apply
the Carleman procedure to the lattice formulation of fluids, as
we describe next.   
  
Besides these similarities, there are also a number of caveats to 
this analogy, which deserves to be discussed in some detail in view 
of some confusing results in the previous literature.

To begin with, it should be appreciated that the Carleman results for the logistic equation
cannot be carried over as-is to the case of the fluid equations, the reason being
that in the latter, the attractor is generally non-local due to spatial coupling due to
advection and diffusion.
This is readily shown already for the simplest case of the Burgers equation 
(pressure-less fluid), by casting this equation in relaxation form: 
\begin{equation}
\partial_t u =  \Delta (u- \frac{Re}{2} \Delta^{-1} \nabla u^2) 
\end{equation}  
$\Delta^{-1}$ being the Green function of the Laplacian 
operator and $\nabla$ the gradient operator.
Since the Green function is nonlocal, it is clear that
the quadratic non-linearity of the Burgers equation, once cast
in relaxation form, leads to a global attractor
\begin{equation}
u^{eq} = \frac{Re}{2} \Delta^{-1} \nabla u^2 
\end{equation}  

This is likely to explain why the Carleman-Burgers procedure does not abide by the 
restriction $Re < \sqrt 2$ advocated in the theorem proposed in  \cite{liu_efficient_2021}.
It is also worth further noting that this theorem was derived from inspection of the unstable attractor of the 
logistic equation a situation which does not relate directly to the 
physics of fluids since even though the local equilibrium $u^{eq}$
can grow in time under a physical instability, the flow configuration $u$
still is attracted to it, as the Green function of the Laplacian
is negative-definite.
We conclude that the restriction $Re < \sqrt 2$ is not relevant
to the Carleman linearization of the fluid equations.
 
\section{Carleman Lattice Boltzmann versus Carleman Navier-Stokes}\label{sec:III}

The satisfactory convergence of the Carleman procedure applied to the
logistic equation points to the Lattice Boltzmann (LB) method as
a natural candidate for a fluid implementation (full details in the original 
publications~\cite{sanavio_lattice_2024,sanavio_three_2024}).
Additionally, the LB method is known for its efficient use of parallel resources, making it a competitive tool for large-scale fluid dynamics simulations~\cite{melchionna_hydrokinetic_2010,falcucci_extreme_2021}.

To illustrate the point, let us remind the basic form of the LB equation:
\begin{equation}
\label{LBE}
f_i(\vec{x}+\vec{c}_i,t+1)-f_i(\vec{x},t) = -\omega (f_i(x,t)-f_i^{eq}(x,t))
\end{equation}
where $f_i(\vec{x},t)$ is the probability of finding a representative fluid particle (discrete
Boltzmann distribution function) at site $\vec{x}$ and time $t$, with a given molecular
velocity $\vec{c}_i$, where $i$ runs over a discrete lattice with suitable symmetry \cite{succi_lattice_2018}.
In the above, the timestep has been set to unity for simplicity.
The left-hand side is the free streaming, whereas the right-hand side encodes
collisions in the form of a relaxation to a local equilibrium $f_i^{eq}$ on 
a timescale $\tau=1/\omega$.
The local equilibrium takes the following form:
\begin{equation}
f_i^{eq} = w_i \bigg{(}\rho + \frac{\vec{J} \cdot \vec{c}_i}{c_s^2}
         + \frac{\rho^{-1} \vec{J}\cdot\vec{J}(\vec{c}_i\cdot \vec{c_i}-c_s^2)}{2c_s^4}\bigg{)}
\end{equation}
where $\rho=\sum_i f_i$ is the fluid density and $\vec{J}=\sum_i f_i \vec{c}_i$ is the 
fluid current density.
The collision operator bears a strong resemblance to the logistic
equation in that not only it is quadratic but it is also local, a property that 
is not and cannot be shared by the Navier-Stokes equations, since it 
derives explicitly from the extra velocity dimensions of phase-space~\cite{benzi_lattice_1992}.

Moreover, since the fluid current is a local collisional invariant, the Carleman 
procedure based on such quantity instead of the 
discrete distribution, is captured {\it exactly} by a second order 
Carleman procedure~\cite{itani_analysis_2022}.   
Unfortunately, this beautiful property is broken by the streaming 
operator, although one may expect that a remnant of this convenient
property may remain also in the space-dependent case.  

Importantly, as first noted in \cite{li_potential_2023}, in the LB method the strength
of the nonlinearity is formally measured by the Mach number instead of the Reynolds number.
The reason is twofold; first, the local equilibria are quadratic in the Mach number, second,
the LB formalism does not involve any second order derivative in space because dissipation
emerges out of local relaxation. The latter is again a distinctive property
of the LB scheme: information always travels along straight lines, constant in space in 
time, regardless of the complexity of the flow field. 
As a result, the expectation is that Carleman-LB (CLB) should meet with 
a much lower nonlinearity barrier as compared to the Carleman-Navier-Stokes (CNS) 
procedure.

In Refs.~\cite{sanavio_lattice_2024} and ~\cite{sanavio_three_2024} we applied CLB and CNS respectively. 
The two formulations led to very different results. 
In both cases, we simulated a two-dimensional Kolmogorov-like flow, defined on a  
$L\times L$ square lattice with the following initial conditions for the lattice site $\vec{x}=(x_1,x_2)$:

\begin{eqnarray}
\rho(\vec{x},0) &=& 1,\\
J_1(\vec{x},0) &=& A_1\cos{k x_2},\\
J_2(\vec{x},0) &=& A_2\cos{k x_1},
\end{eqnarray}

\noindent where $J_a$, with $a=1,2$, is the fluid current density in the two dimensional 
plane and the wavenumber is set to $k=2\pi/L$. 
We simulated the dynamics on a $32\times32$ grid with periodic boundary conditions 
and we compared the solutions for the macroscopic quantities $\rho,\vec{J}$ 
obtained by the CLB and by the CNS procedures. 
By modifying the value of the amplitudes $A_1,A_2$, we explored different 
regimes of nonlinearity. 

In fact, if either $A_1$ or $A_2$ is set to zero, the
nonlinear terms of the fluid equation vanish by construction.
In this case, the current $J_a$ decays exponentially in time as

\begin{equation}\label{eq:linear}
J_{1,2}(\vec{x},t) = J_{1,2}(\vec{x},0)e^{-\nu k^2 t},
\end{equation}

\noindent with $\nu$ being the kinematic viscosity.

\begin{figure}
    \centering
    \includegraphics[scale = 0.4]{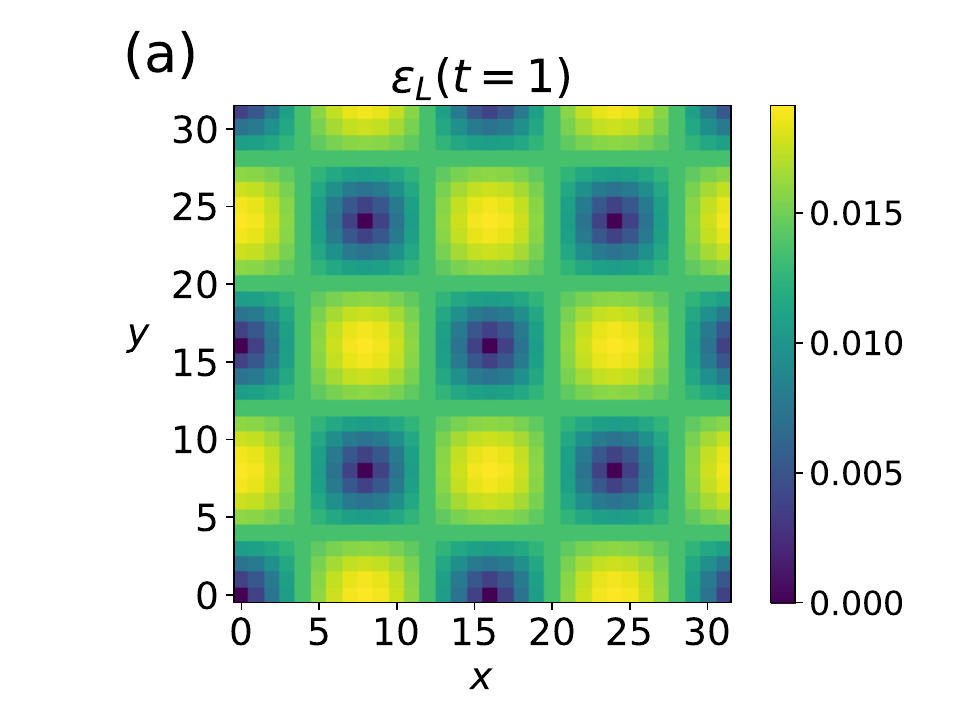}
    \includegraphics[scale = 0.4]{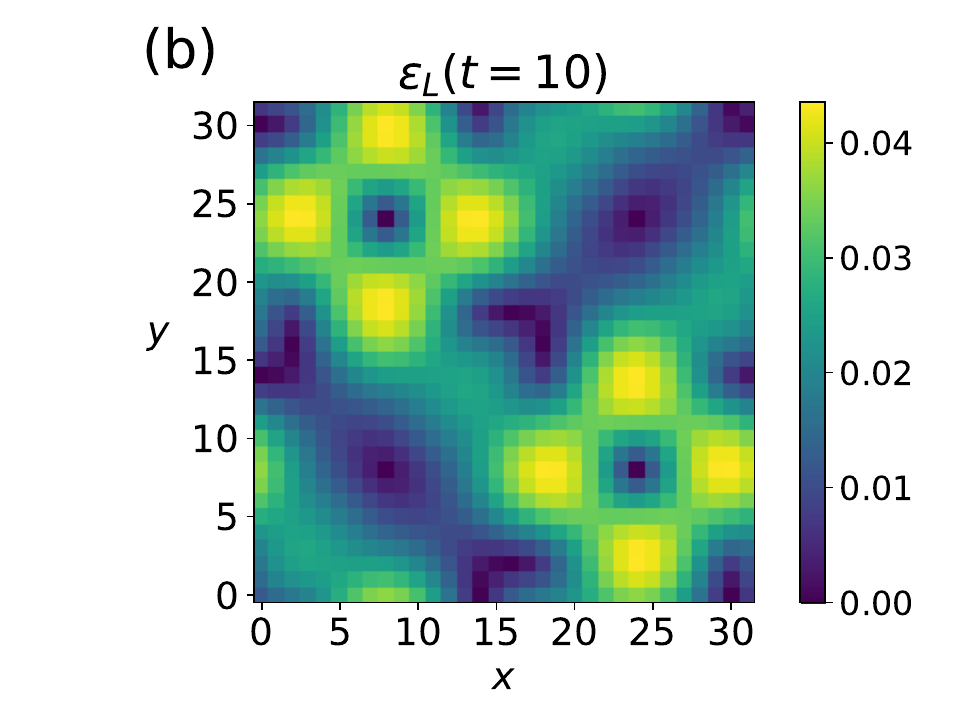}
    \caption{\label{fig:carlemanLB} 
The relative error~\eqref{eq:relative_error} of the CLB method truncated at second order on each site of \
a $32\times32$ lattice, for a Kolmogorov-like flow with initial conditions $A_1=A_2=0.1$, $\nu=1/6.$}
\end{figure}

We set the viscosity to $\nu=1/6$, the amplitude to  $U=A_1=A_2=0.1$ and $L=32$, yielding $Re = \frac{UL}{\nu} \sim 20$. 
The CLB numerical simulations provided three-digit convergence
up to about $100$ timesteps, even with the lowest Carleman truncation, namely second order. 
This is shown in Fig.~\ref{fig:carlemanLB}, where we plot the relative error $\epsilon_L$, defined as

\begin{equation}\label{eq:relative_error}
\epsilon_L = \sum_{a=1,2}|\frac{J_a^{\text{LBM}}-J_a^{\text{CLB}_2}}{J_a^{\text{LBM}}}|,
\end{equation}

\noindent which accounts for the difference between the values of $\vec{J}$ obtained 
with direct numerical simulation of the LBM and the value obtained by Carleman-LB linearization. 
The relative error is plotted at two different times, $t=1$ and $t=10$, and it is found
to stabilize at larger times. 

% -----------------------------
% -----------------------------

\begin{figure}
    \centering
    \includegraphics[scale = 0.4]{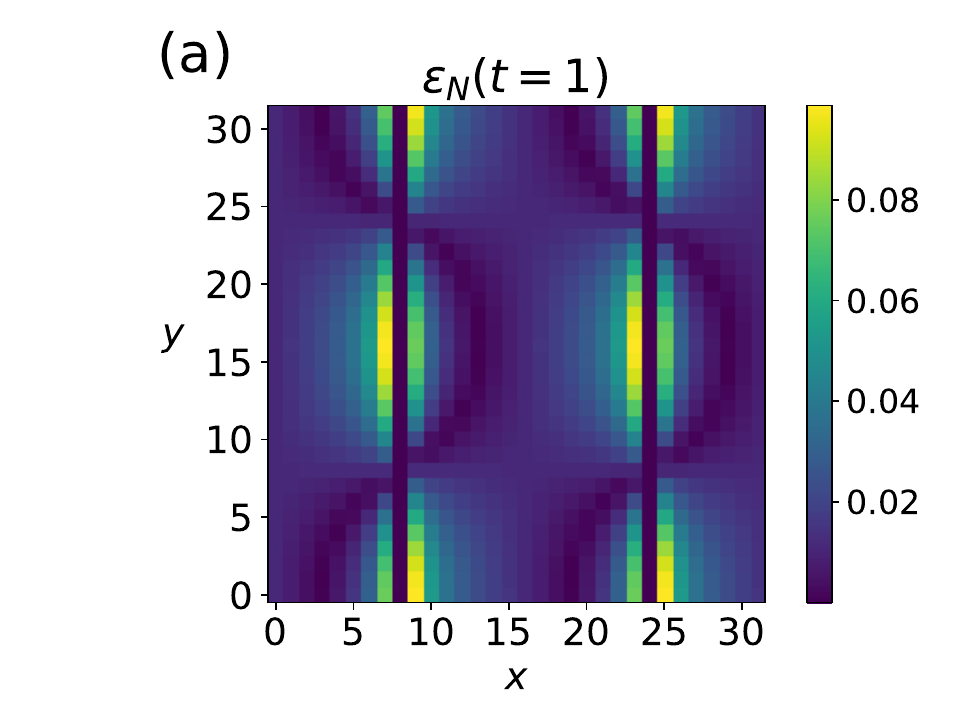}
    \includegraphics[scale = 0.4]{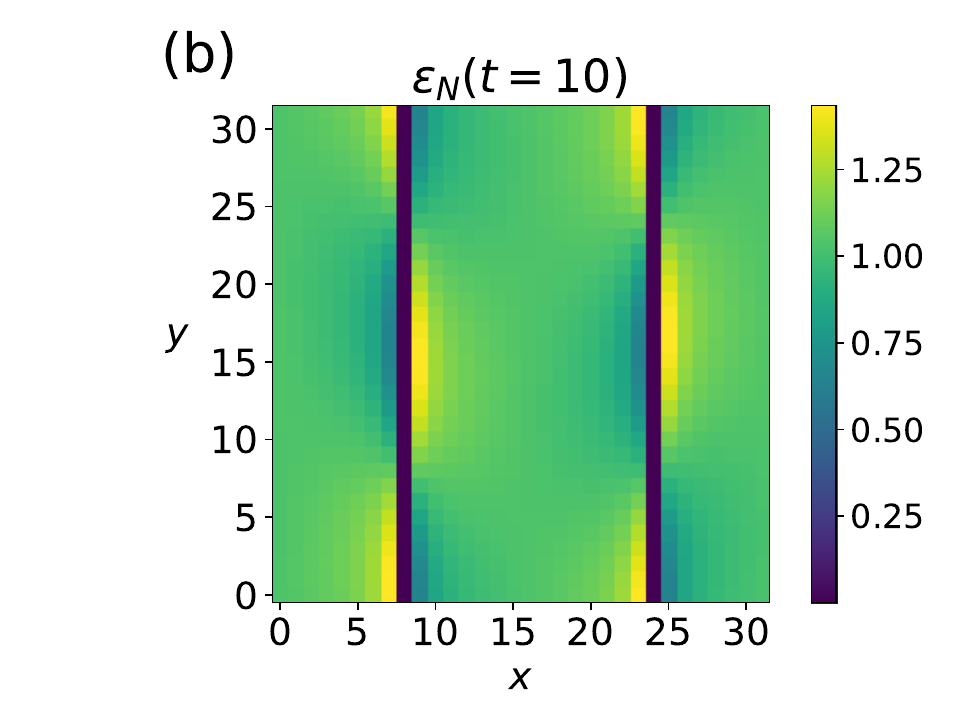}
    \caption{\label{fig:carlemanNS} 
The relative error~\eqref{eq:relative_error} of the CNS method truncated at second order on each site of a $32\times32$ lattice, for a Kolmogorov-like flow with initial conditions $A_1=A_2=0.1$, $\nu=1/6.$}
\end{figure}

By contrast, by using CNS, we have not been able to preserve a similar accuracy beyond
a time horizon of a few time steps, (with $dt=0.01$), even at fourth order in the Carleman truncation. 
The relative error $\epsilon_N$ defined in a similar manner as 
Eq.~\eqref{eq:relative_error} for the CNS method is plotted in 
Fig.~\ref{fig:carlemanNS} for the same times $t=1$ and $t=10$.
As one can appreciate, at time $t=10$ the CNS error is about thirty 
times larger than the CLB error. 

% -----------------------------
% -----------------------------

Although much more validation work is needed to put these preliminary 
observations on firm ground, these numerical results are fairly consistent 
with the theoretical perspective discussed earlier on in this paper, namely that
LB should be much more amenable to Carleman linearization than the Navier-Stokes equations. 

As commented above, the stream-relax structure of the LB method brings about
very substantial advantages, primarily a Mach-controlled nonlinearity as 
well as unitarity, to be discussed shortly.

However, the excellent properties of LB come at a steep price, non-locality, as we are going
to comment in the next section.

\subsection{Nonlocality}

A major shortcoming  of the Carleman procedure is the exponential
growth of the number of variables with the order of the Carleman expansion. 
Let $f_i$, $i=1,b$ the number of CLB variables at level $1$  for a LB scheme with
$b$ discrete velocities $\vec{c}_i$, with $b=9$ in two spatial dimensions 
and $b=19$ in three.
In a local scheme, the Carleman variables at level two are 
$f_{ij} = f_i f_j$, hence $b(b+1)/2$ components, while at level $k$ the 
total number of CLB variables is:

\begin{equation}
N_{CLB}(k) = b + b(b+1)/2 + \dots  +b(b+1)\dots(b+k)/k! \sim b^k.
\end{equation}

As an example, the two cases $b=9$ and $b=19$ up to $k=10$ are 
reported in table~\ref{tab:table}, along with the corresponding 
number of required qubits.

\begin{table}
\centering
\begin{tabular}{||l||r|r||r|r||} 
\hline
$k$     &       $N_{9}$    &    $Q_{9}$      &    $N_{19}$     &      $Q_{19}$\\ [0.5ex] 
\hline
\hline
1      &      9    &     4     &       19      &    5\\
2      &     54    &     6     &      209     &    8\\
3      &     219   &     8     &      1539    &    11\\
4      &     714   &     10    &      8854    &    14\\
5      &     2001  &     11    &      42503   &    16\\
6      &     5004  &     13    &      177099  &    18\\
7      &    11439  &     14    &      657799  &    20\\
8      &    24309  &     15    &      220074  &    22\\
9      &    48619  &     16    &      6906899 &    24\\
10     &    92377  &     17    &      20030009&    26\\ [1ex] 
 \hline
\end{tabular}
\caption{Number of Carleman variables and corresponding number of qubits 
for the cases $b=9$ and $b=19$, respectively.
}
\label{tab:table}
\end{table}

As one can see, the number of required qubits grows less than linearly
with the number of Carleman levels, and remains fairly moderate up to
level 10, which is taken here as an upper bound for satisfactory 
Carleman convergence.

Clearly, locality is key to this favourable scaling.
If, on the other hand, the Carleman scheme involves Carleman variables sitting
at different lattice sites, say $f_{ij}(\vec{x}_i,\vec{x}_j)$, then the number of 
Carleman variables after $T$ timesteps scales like $(Tb)^k$ instead of $b^k$.
Since $T$ scales at least linearly with the linear size of the domain, even moderate
grid sizes, say $G \sim 10^6$ in three spatial dimensions, quickly lead to unviable schemes.

Unfortunately, CLB appears to be inherently nonlocal, since, by construction, different populations,
say $f_i(\vec{x},t)$ and $f_j(\vec{x},t)$ at the same location $\vec{x}$ at time $t$, free-stream
to different locations, $\vec{x}_i = \vec{x}+\vec{c}_i$ and 
$\vec{x}_j=\vec{x}+\vec{c}_j$ at time $t+1$, thus generating
a non-local pair $f_{ij}(\vec{x}_i,\vec{x}_j,t+1)$.

In this respect, CNS fares better because, by using conservative differencing in space,
locality can be preserved at all Carleman levels \cite{sanavio_three_2024}.
However, as discussed above, the CNS method shows too poor convergence 
to offer a viable solution for the quantum simulation of fluids.

Given such a state of affairs, we set out to explore a third avenue, the
Carleman-Grad (CG) approach, in the hope it may combine the good convergence 
of CLB with the locality of CNS.

\section{Grad's generalized hydrodynamics}

The basic idea of Grad generalized hydrodynamics is to take progressive
moments of the Boltzmann probability distribution and inspect 
the resulting open hierarchy of hyperbolic PDEs. 
Grad's thirteen-moment method is a technique to derive
solutions of the Boltzmann equation in fluid-like
regimes beyond the realm of Navier-Stokes hydrodynamics \cite{succi_lattice_2018,grad_statistical_1952}.
Formally, it amounts to projecting the Boltzmann equation
upon a suitable set of basis functions in velocity (momentum)
space, typically tensor Hermite polynomials for the case of
cartesian coordinates. The expansion is truncated to third
order terms, delivering the equations of motion for the fluid density $\rho= \int f d^3v$, 
the fluid current $J_a=\int f v_a d^3v$, the momentum flux tensor 
$P_{ab}=\int f v_a v_b d^3 v$, and the energy flux 
$Q_{abc}=\int v_av_bv_c d^3 v $, with latin indexes running over the spatial dimensions.  

This delivers the following set of hyperbolic equations:

\begin{eqnarray}
\partial_t\rho +\partial_aJ_a&=& 0 \label{eq:mass}\\
\partial_tJ_a +\partial_bP_{ab}&=& 0 \label{eq:current}\\
\partial_tP_{ab} +\partial_cQ_{abc}&=&-\omega(P_{ab}-P_{ab}^{\text{eq}}),\label{eq:momentum}
\end{eqnarray}

This system of equations generalizes the
Navier-Stokes equations (NSE) in that all physical signals travel
at finite speed, as opposed to the parabolic character of
the dissipative term of the NSE.

Like for LB, the NSE are recovered in the limit of weak departure from local
equilibrium. Under such limit, the third equation can be closed by
assuming adiabatic relaxation of $P_{ab}$ to its local equilibrium, namely:
\begin{equation}
P_{ab} \sim P_{ab}^{\text{eq}} -\tau \partial_c Q_{abc}^{\text{eq}}
\end{equation}
where $\tau=1/\omega$ and the equilibrium tensors read as follows:

\begin{eqnarray}\label{eq:equilibrium}
P_{ab}^{\text{eq}}&=& \frac{J_aJ_b}{\rho}+c_s^2\rho\delta_{ab} \nonumber\\
Q_{abc}^{\text{eq}}&=& c_s^2(J_a\delta_{bc}+J_b\delta_{ac}+J_c\delta_{ab}),
\end{eqnarray}
\noindent where $c_s$ is the speed of sound.

Note that in the latter we have neglected cubic nonlinearities by 
assuming low-Mach number conditions.

The appeal of the Grad formulation rests with  its hyperbolic character, which reflects
in the conservative nature of the equations, hence the locality of the discretized scheme.
In addition, the number of native variables is  $1+3+6=10$, half than CLB.

Despite its conceptual and mathematical elegance, Grad's method 
suffers a major drawback: lack of realizability, i.e.  it 
does not ensure the positive definiteness of the 
truncated distribution function,  often reflecting instability issues. 
Regularization techniques have been developed over the years which permit 
to describe several far-from-equilibrium transport phenomena 
as they typically arise in the context of rarefied gas dynamics \cite{struchtrup_regularization_2003}.
Hence, the regularized CG scheme could form the basis to simulate
these problems on quantum computers, offering an alternative to expensive 
Monte Carlo procedures.

\subsection{Grad versus Lattice Boltzmann}

It is worth recapping the points of commonality and departure between Grad
and Lattice Boltzmann.

Commonalities: i) they are both first order in space and time,  
ii) dissipation emerges from local relaxation with no need of second order
spatial derivatives, iii) the nonlinearity is local, hence formally measured by the Mach
number instead of the Reynolds number.

Point of departure: they differ in the structure of spatial transport.
The key point is that, at variance with the fluid equations, the LB information invariably
travels along straight characteristics, $d\vec{x}_i = \vec{c}_i dt$, constants in space and time.
As a result, the left-hand side of the LB equation is exact,  a property that
lies at the heart of the success of the LB method for classical fluids.
In the CLB context, this means that the streaming operator remains unitary also upon lattice
discretization. Indeed, the collisionless LB update ($\omega=0$) is simply the identity
$f_i(\vec{x}+\vec{c}_i,t+1)=f_i(\vec{x},t)$, a property which can be shown to hold at all
Carleman levels.
In the Grad method, however, fluid transport is implemented via first-order spatial 
derivatives of the generalized currents. Hence, besides being subject to discretization 
and round-off errors, such discrete derivatives do not preserve unitarity under 
Euler time marching. 
In this respect, the Grad method aligns with the Navier-Stokes equations.

With all these plus and minuses factored in, the question remains as to
whether the Grad formulation can benefit from the excellent convergence
of CLB, while sharing the locality of the CNS formulation.
In the following we investigate such question by first formulating the 
Carleman-Grad procedure and then by performing numerical simulations.

\section{The Carleman-Grad procedure}
 
The first order Carleman step is to set the nonlinear terms
$J_{ab}=0$ and $J_{abc}=0$. 
Further Carleman steps require knowledge of $J_{ab}$ to 
compute $P_{ab}^{\text{eq}}$. The dynamic equation for $J_{ab}$ can be 
obtained the same way as for CNS, with the advantage of more locality 
since no Laplacian operators are involved. 

In order to apply the Carleman linearization and conserve the local form of the equations at all Carleman 
orders, we proceed as follows. 
First we use the approximation valid for a weakly-compressible flow $\rho^{-1}\approx 2-\rho$, in 
order to express all the equations~\eqref{eq:mass}--\eqref{eq:momentum} in polynomial form.
We collect all the local variables in a vector $V(\vec{x},t)=(\rho,J_a,P_{ab})$, 
with $a,b$ running over the spatial dimensions, and rewrite 
Eqs.~\eqref{eq:mass}--\eqref{eq:momentum} in matrix form as:

\begin{eqnarray}
\partial_tV(\vec{x},t)&=& \hat{L}_1 V(\vec{x},t)+
\hat{L}_2 V^{(2)}(\vec{x},t)+\hat{L}_3 V^{(3)}(\vec{x},t) \label{eq:Carleman},
\end{eqnarray}

\noindent where the matrices $\hat{L}_j$, with $j=1,2,3$ multiply the linear, quadratic and cubic terms respectively. 
Note that the matrices $\hat{L}_2$ and $\hat{L}_3$ are very sparse, since the non-zero 
terms come only from the equilibrium tensor $P_{ab}^{\text{eq}}$ \eqref{eq:equilibrium}. 
Their explicit form can be found in the Appendix.

Here we explicitly apply the Carleman procedure up to third order, thus ignoring the terms $ V^{(k)}$, with $k>3$, but the generalization is straightforward for higher truncation orders.
We discretize time via the Euler forward scheme, yielding the following relation for each local vector:

\begin{eqnarray}\label{eq:CarlemanEuler}
V_a(\vec{x},t+1)&=& A_{ab}V_b(\vec{x},t)+B_{abc} V^{(2)}_{bc}(\vec{x},t)
+C_{abcd} V^{(3)}_{bcd}(\vec{x},t),
\end{eqnarray}

\noindent where we have defined $A=\mathds{1}+dt \hat{L}_1$, $B=dt \hat{L}_2$ and $C=dt \hat{L}_3$.
The second and third-order Carleman variables are the tensor products of the local vectors,
\begin{equation}\label{eq:Carleman_2nd}
V^{(2)}(\vec{x},t) = V(\vec{x},t)\otimes V(\vec{x},t),
\end{equation}

\noindent and their evolution can be obtained by tensor product of Eq.~\eqref{eq:CarlemanEuler} as

\begin{eqnarray}\label{eq:Carleman_evolution}
V^{(2)}(\vec{x},t+1)&\equiv& V(\vec{x},t+1)\otimes V(\vec{x},t+1)\nonumber\\
%&=&[A_{ac}V_c(\vec{x},t)+B_{acd} V^{(2)}_{cd}(\vec{x},t)+C_{acde} V^{(3)}_{cde}(\vec{x},t)]^{\otimes 2}\nonumber\\
%&=&(A\otimes A) V(\vec{x},t)\otimes V(\vec{x},t)+B\otimes A V^{(2)}(\vec{x},t)\otimes V(\vec{x},t)\nonumber\\
%&&+A\otimes B V(\vec{x},t)\otimes V^{(2)}(\vec{x},t)\nonumber\\
%&&B_{acd}\otimes B_{bef} V^{(2)}_{acd}(\vec{x},t)\otimes V^{(2)}_{bef}(\vec{x},t)+\dots\nonumber\\
&=&(A\otimes A) V^{(2)}(\vec{x},t)+(B\otimes A+A\otimes B) V^{(3)}(\vec{x},t),\nonumber\\
V^{(3)}(\vec{x},t+1) &=& (A\otimes A\otimes A)V^{(3)}(\vec{x},t)
\end{eqnarray}

\noindent and all the higher orders terms are neglected.

\section{Carleman-Grad: numerical results}

\begin{figure}
    \centering
    \includegraphics[width=0.3\textwidth]{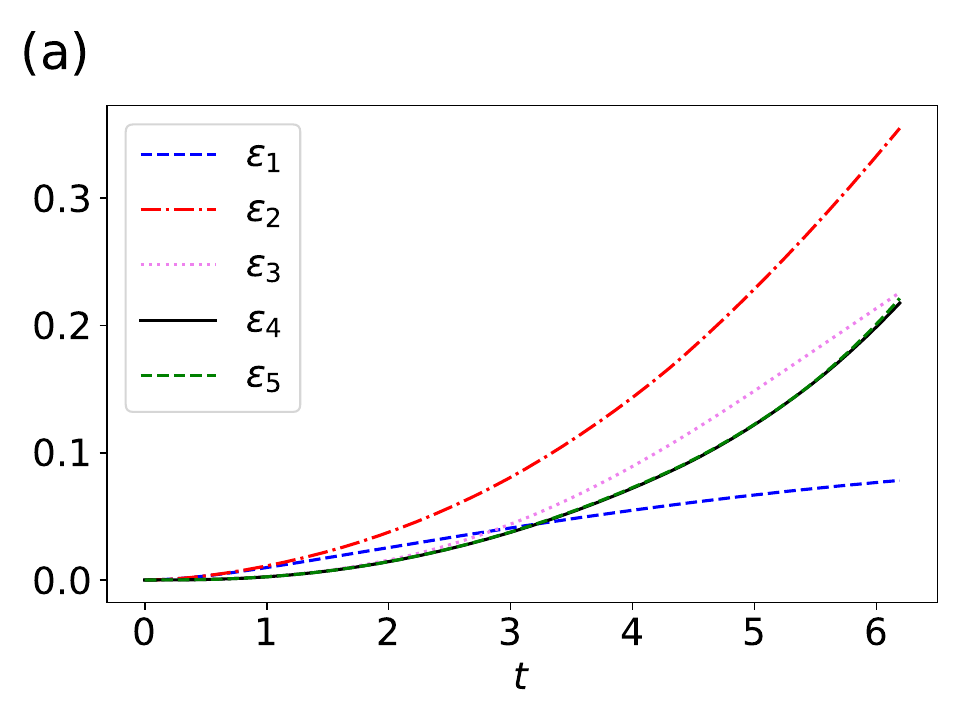}
    \includegraphics[width=0.3\textwidth]{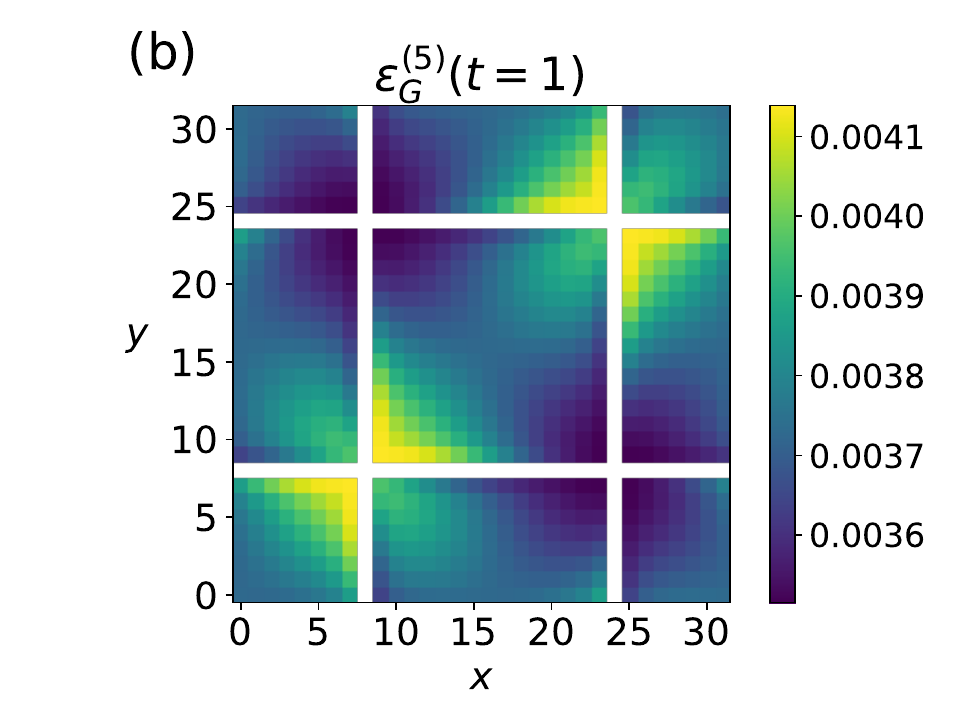}
    \includegraphics[width=0.3\textwidth]{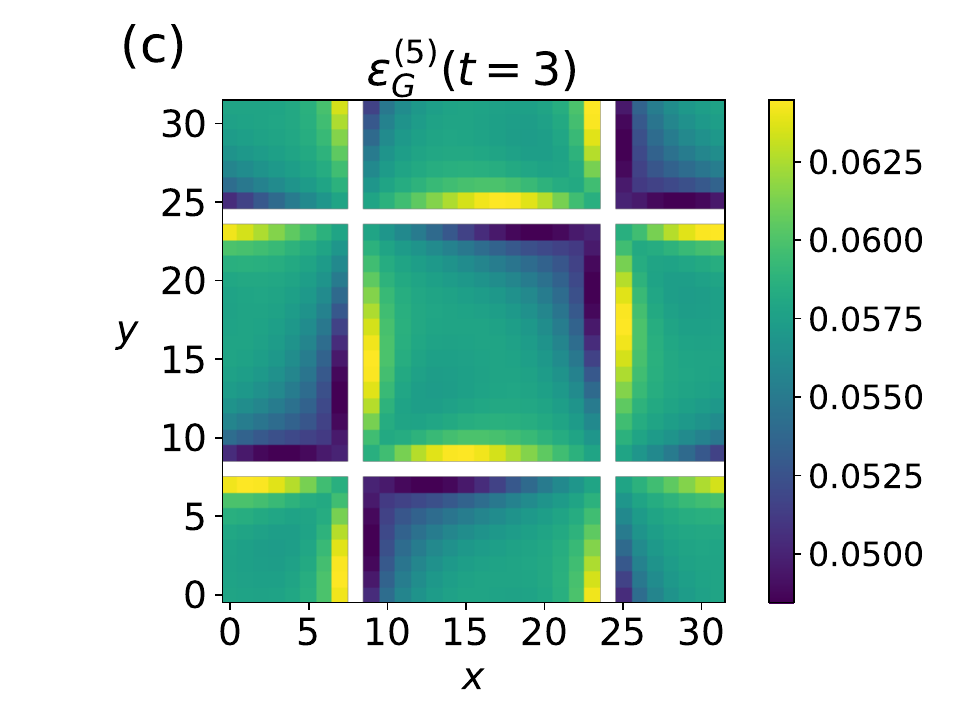}
    \caption{\label{fig:carlemanGrad} 
(a)The time evolution of the relative errors $\epsilon_k$ of the CG method truncated at order $k$ averaged over the lattice gridpoints, for a Kolmogorov-like flow with initial conditions $A_1=A_2=0.1$, and $\omega=2$. 
(b) The relative error of the CG method with $k=5$ plotted at each lattice site at time 
$t=1$. (c) The relative error of the CG method with $k=5$ plotted for each lattice site at time $t=3$. 
The points on the white bands are removed from the plot since $J$ was initialized to $0$.}
\end{figure}

Next, we show the results obtained with the CG method up to 
order $5$ for the simulation of the Kolmogorov-like flow described in Sec.~\ref{sec:III}.
In Fig.~\ref{fig:carlemanGrad}(a) we show the mean value $\epsilon_k$  of the relative error $\epsilon_G^{(k)}$, for the CG linearized system truncated at order $k$, obtained by averaging over the 
entire set of gridpoints. 
In Figs.~\ref{fig:carlemanGrad}(b) and~\ref{fig:carlemanGrad}(c) we show the relative error 
at each site of the lattice for $k=5$ at $t=1$ and $t=3$, respectively. 
The latter is chosen since the first and the fifth order approximation yield the same value of $\epsilon_k$.

% -----------------------------
% -----------------------------

\begin{figure}
    \centering
    \includegraphics[width=0.9\textwidth]{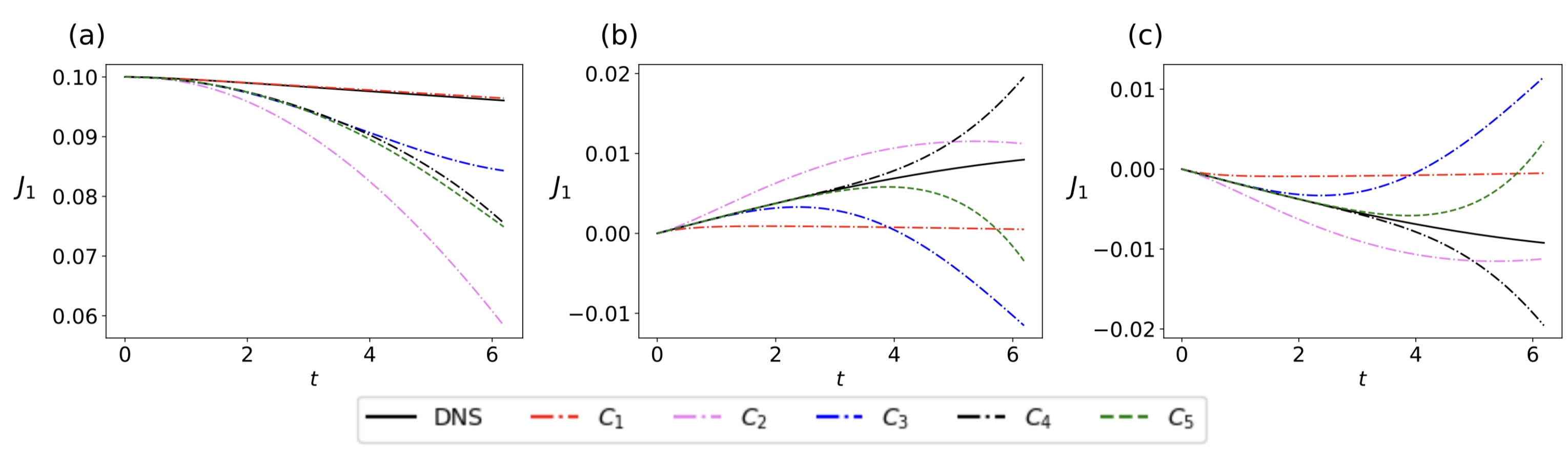}
    \caption{\label{fig:carlemanGradloc} 
The time evolution of the current density $J$ of the CG method truncated at different orders $k$ for specific representative lattice gridpoints, for a Kolmogorov-like flow with initial conditions $A_1=A_2=0.1$, $\omega=2$ and $dt=0.01$. The speed of sound is set $c_s=\frac{1}{\sqrt{3}}$. The chosen points are $\vec{x}_a=(0,0)$, $\vec{x}_b=(8,0)$ and $\vec{x}_c=(8,16)$ in (a), (b) and (c) respectively.}
\end{figure}

Figures~\ref{fig:carlemanGradloc}(a)-(c) compare the values of  $J_1$
​obtained from CG dynamics at different orders with those obtained from direct numerical simulation (DNS) 
of Grad's equations at three lattice sites: $\vec{x}=(0,0), \vec{x}=(8,0)$, and $\vec{x}=(8,16)$, respectively.

% -----------------------------
% -----------------------------

Each curve exhibits a distinct behaviour and convergence times. 
At $\vec{x}=(0,0)$,  the dominant dynamics leads to a current decay, hence 
the first-order Carleman truncation best captures the dynamics at later times, even though 
higher-order approximations are initially more accurate.  
In contrast, at points $\vec{x}=(8,0)$ and $\vec{x}=(8,16)$, the convergence times for 
the fourth and fifth-order CG dynamics are similar, mirroring the behaviour observed 
in Fig.~\ref{fig:carleman_logistic}(a) for the growing logistic equation. 
We speculate that such similarity may point to the existence of a suitable transformation 
of the Carleman variables more conducive to a convergent solution.
For instance, one could write a dual Navier-Stokes equation for $u^{eq}-u$, where $u^{eq}$ 
is the nonlocal equilibrium discussed in the early part of this paper, a possibility that remains
to be explored in the future.

\section{Time marching, unitarity, and depth of the quantum circuit}

In this section, we analyze the time marching scheme to solve 
the Carleman linearized evolution problem: 

\begin{equation}
\label{CONTI}
\frac{dV^{(k)}}{dt} = C_k V^{(k)}, \;k=1,K
\end{equation}
where $V^{(k)}$ is the array of Carleman variables at level $k$ of the procedure
and $C_k$ is the corresponding Carleman matrix.

\subsection{Single-step telescopic Euler scheme}

Time marching is a notoriously hard issue on quantum computers, due to
the measurement problem: the updated state at time $t+dt$ cannot be copied
into the initial state for the next time step, since this would
require a full-state tomography of complexity $\mathcal{O}(4^Q)$.

This problem can be circumvented by adopting a 
single-step telescopic forward Euler scheme:

\begin{equation}
\label{EULER}
V^{(k)}(t) = (\mathds{1}-C_k \Delta t)^T V^{(k)}(0)
\end{equation}

where $t = T \Delta t$ and $\mathds{1}$ is the identity matrix.

The above scheme propagates the initial condition at $V^{(k)}(0)$ 
to the final state at $V^{(k)}(t)$ in a single large step of size $t$ with the same
accuracy of $T$ timesteps of size $dt=t/T$, so as to avoid any intermediate measurement. 
Of course, this does not come for free.
In fact, if the $C_k$ matrix has a bandwidth $B_k$, the telescopic
propagator has bandwidth $TB_k$, hence, for any $T$ comparable with the linear
size of the domain, this amounts to a full matrix even in the case of a sparse matrix $C_k$.   

We further note that the matrix $C_k$ consists of a unitary streaming component
and a non-unitary one, in the form of a local collision-relaxation operator
for CLB and a Laplacian operator for CNS.
 
In the CLB scheme we chose to represent the collision matrix as a weighted superposition
of two unitaries, according to the procedure first discussed in \cite{mezzacapo_quantum_2015}. 
This has the merit of simplicity, but it comes at a price.
Indeed, by writing the collision matrix as a weighted sum of two unitaries, the corresponding 
single-step quantum circuit update fails with a non-zero probability $p_f(\gamma)>0$. 
Hence the failure probability over a sequence of $T$ timesteps is $p_f^T$, meaning that
in order to keep the  multistep failure probability well below 1, optimal 
values of $\gamma$ must be found such that $$p_f(\gamma) < 1/T.$$
The best value we have found to date for CLB is $p_f \sim 0.09$, which allows us to
propagate about 10 steps with an acceptance ratio of about $1/3$.
Clearly, smaller values of $p_f$ are needed to advance over hundreds or thousands
of timesteps.

We note, however, that such a restriction could be sidestepped by turning
to quantum linear algebra solvers, to which we shall return in the final part of this paper.  

It is also important to notice that, even in the case where the continuum 
problem, Eq. (\ref{CONTI}), is unitary, the Euler scheme, Eq. (\ref{EULER}),
is generally not. However, contrary to CNS, CLB  does preserve unitarity of the 
streaming operator.
Finally, we inspect the depth of the quantum CLB algorithm, namely the number of two 
qubits gates required to implement the CLB  telescopic operator.
 
To date, we have not been able to find any CLB matrix representation
leading to a circuit depth below the upper-bound value $4^Q$.
Indeed, projection over a tensorial basis of native quantum gates delivers 
an exponentially long series, whose coefficients do not show any significant
decay in amplitude over the entire basis set.
In this respect, the CLB matrix appears as irreducible as a generic unitary,
thus compromising the viability of CLB scheme, despite its excellent convergence properties.
If one could find a reducible representation, then CLB would turn up into a serious candidate 
for the quantum simulation of fluids, but for the moment it is not.
On the other hand, CNS offers no alternative, due to its very poor Carleman convergence
and CG sits in a sort of intermediate position. 

Next, we present our perspectives for a quantum implementation of CG.  

\subsection{Perspectives for a Carleman-Grad quantum circuit based on quantum linear solvers}

The local formalism introduced with Eqs.~\eqref{eq:Carleman_evolution} for the Grad-method
leads to a matrix characterized by low sparsity. 
This property is crucial for the implementation of an efficient quantum circuit reproducing the behaviour of the Carleman matrix~\cite{berry_efficient_2007,berry_black-box_2023}. 

In fact, considering the simple case $k=2$, the evolution of the global Carleman array
\begin{equation}
\vec{V}=\left(V(\vec{x}_1),\dots,V(\vec{x}_N),V^{(2)}(\vec{x}_1),\dots V^{(2)}(\vec{x}_N)\right)
\end{equation}

\noindent for a lattice with $N$ gridpoints and a single time step $dt$, writes as follows:

\begin{equation}\label{eq:Carleman_2}
\vec{V}(t+dt)=
\begin{pmatrix}
\hat{A} & \hat{B}\\
0 & \hat{A}\otimes \hat{A}
\end{pmatrix}
\vec{V}(t),
\end{equation}

\noindent where the matrices $\hat{A},\hat{B}$ are given in Appendix~\ref{app:I}. 

In order to solve the linear system of equations~\eqref{CONTI} we refer to two quantum linear solvers. 

The first is the algorithm proposed by Harrow, Hassidim, and Lloyd 
(HHL)~\cite{harrow_quantum_2009}, featuring a 
computational complexity $\mathcal{O}(\log[g^kN]s^2\kappa^2/\epsilon)$, where $\kappa$ 
is the condition number  of $C_k$ and $\epsilon$ is the desired precision. 
The second, is the quantum solver proposed by Childs, Kothari, and Somma (CKS)~\cite{childs_quantum_2017} 
which improves the HHL algorithm by reducing the computational complexity to $\mathcal{O}(\log[g^kN/\epsilon]\kappa)$.

Although the CKS algorithm consistently improves the dependency on $\epsilon$, its implementation 
necessitates a black-box oracle Hamiltonian to replicate the matrix $C_k$, thereby severely 
hampering its practical application. 
By contrast, the HHL algorithm is more readily applicable, at least in principle, to the present investigations. 

The sparsity $s$ of the Carleman matrix $C_k$ of Eq.~\eqref{CONTI} for the local Grad 
system scales $6^k$, to be compared with 
the size of the matrix $C_k$ itself, namely $6^kN\times 6^kN$.
Note that the $\log N$ quantum advantage of both HHL and CKS algorithms 
can be totally spoiled by an unfavourable dependency of the condition number $\kappa(N)$~\cite{montanaro_quantum_2016} with the number of sites. 

\begin{figure}
    \centering
    \includegraphics[scale = 0.4]{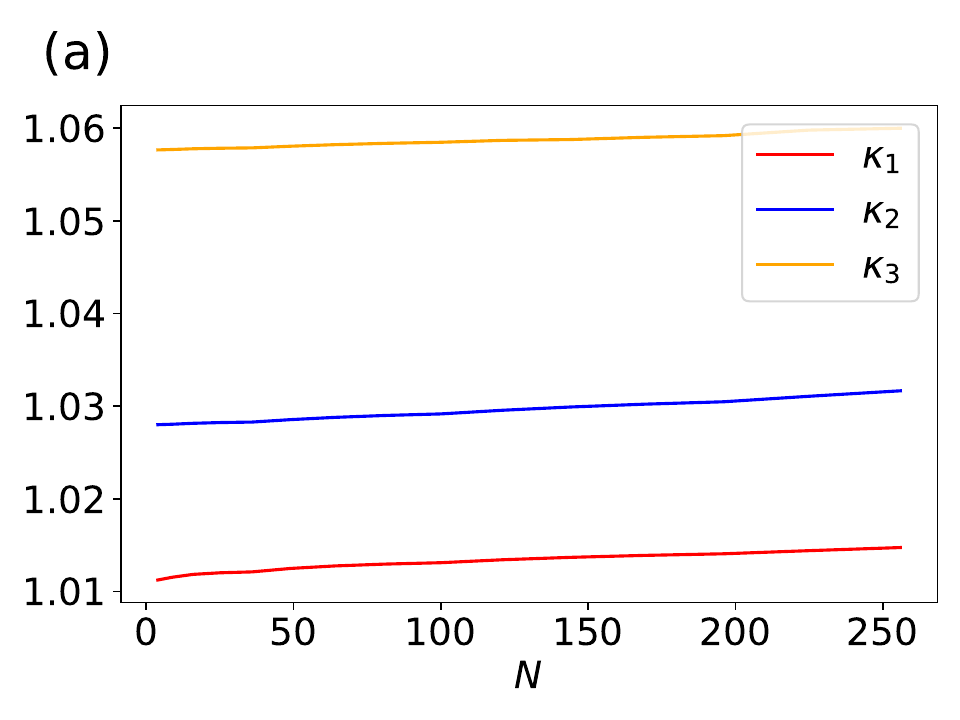}
    \includegraphics[scale = 0.4]{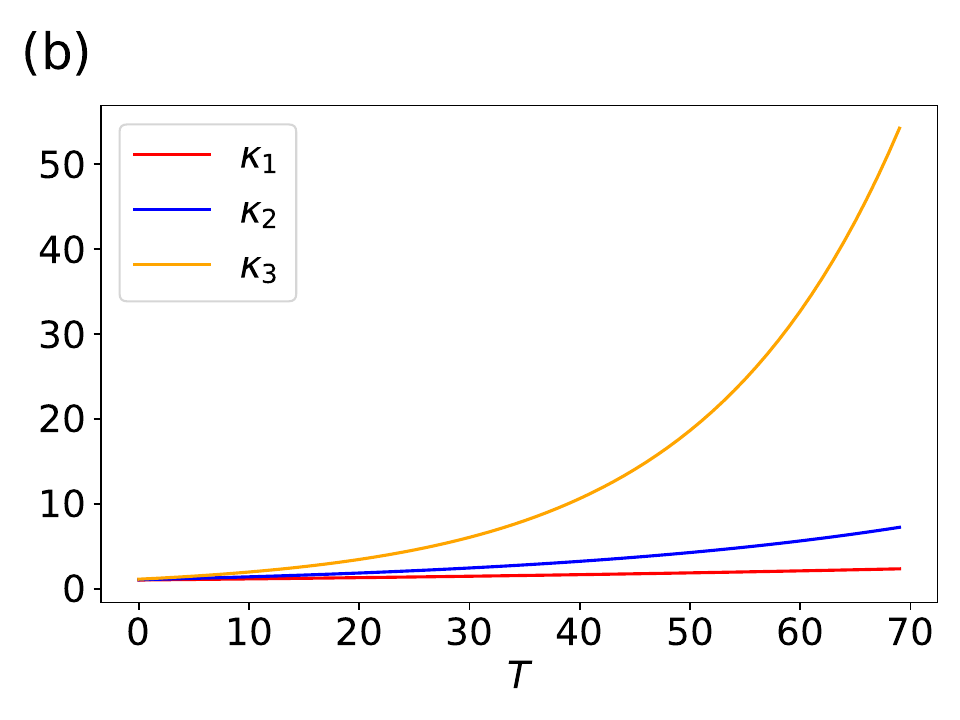}
    \caption{\label{fig:condition} 
(a)The condition number $\kappa$ of the Carleman matrix $C_k$ with truncation order $k=1,2$ and $3$ varying the number of lattice sites. (b) The condition number $\kappa$ of the Carleman matrix $C_k$ with truncation order $k=1,2$ and $3$ as a function of the number of timesteps. The parameters of the CG matrix have been set to $\omega=2$ and $dt=0.01$.}
\end{figure}

To investigate this point, we have explicitly calculated the value of $\kappa$ as 
a function of $N$ for the case $\omega=2$ and found a
sub-linear dependence on $N$, as reported in Fig.~\ref{fig:condition}(a). 

% -----------------------------
% -----------------------------

In order to obtain the solution at a given time $t$, we need to apply the Carleman matrix to the initial 
vector state $T=t/dt$ times, with $\vec{V}(t)=C^{T}\vec{V}(0)$. 
In Fig.~\ref{fig:condition}(b) we plot the condition number as a function of  the 
number of time steps, $\kappa(T),$ for the case $\omega=2, dt =0.01$ and 
show that it exhibits a polynomial dependence on $T$. 
As a result, the trend of $\kappa$ with both $N$ and $T$ justifies some 
optimism for future quantum implementations of the CG method, provided
convergence and circuit depth can be kept under control.

To date, we have not been able to address the other key requirement of an 
efficient quantum implementation of the Carleman matrix: block-diagonality~\cite{berry_efficient_2007}. 
This leaves us again with a $4^Q$ depth of the CG quantum circuit, like for CLB.
 
Further investigations are needed to address this crucial point. 

\section{Conclusion and outlook}

In this paper, we have investigated the application of the Carleman procedure 
to the Grad's formulation of generalized hydrodynamics. 
This analysis was motivated by previous works~\cite{sanavio_lattice_2024,sanavio_three_2024}, 
where we analysed the Carleman LB and the Carleman NS methods, respectively. 
The first (CLB) shows excellent convergence between the linearised and the classical 
solution for a large  number of time steps, in the order of hundreds. 
However, its inherent nonlocality generates a number of Carleman variables which 
grows exponentially with the truncation order.
As a result, the number of two-qubit quantum gates of the CLB circuit 
scales exponentially with the number of qubits. 

The second method (CNS) can be written using a local formalism, thus allowing the number of Carleman variables 
to scale linearly with the number of lattice sites. However, the simulations show that 
a satisfactory convergence between the Carleman and the classical solution can be achieved only 
for a few timesteps. This is due to a number of reasons, not least the lack of 
unitarity of the time-marching scheme. 

As expected, Carleman Grad sits somewhere in between. 
Our results show good convergence of the linearization method up to 
about thirty timesteps at $k=5$-th order of truncation, and the analysis of the 
quantum complexity points to potentially viable implementations using
quantum linear solvers.
However, the convergence horizon is still at least one order of magnitude
too short for practical fluid-dynamic purposes.
%(we remind that with initial velocity $|\vec{u}|=0.1$ and a timestep $dt=0.1$ ,
%$t=320$ timesteps are necessary to cover a full cycle around a $32 \times 32$ lattice).

Many directions for future work remain to be explored, such as the 
search of more efficient decompositions of the CLB matrix, as well
as improvements of the convergence of the CG procedure by resorting to more advanced 
space and time discretization techniques, along with reduced sparse-matrix representations. 
Yet another direction is Carleman linearization based on orthogonal
polynomials, such as Tsebitchev, which may significantly improve convergence.  
Only the future can tell which one will take us closes to a quantum algorithm for fluid flows.  

\section*{Acknowledgements}

The authors have benefited from valuable discussions 
with many colleagues, particularly
M. Lauricella, R. Scatamacchia.
The authors acknowledge financial support from the Italian National
Centre for HPC, Big Data and Quantum Computing (CN00000013). 

The authors have no conflicts to disclose.
The data that support the findings of this study are available from the corresponding author upon reasonable request.

\nocite{}
\bibliographystyle{unsrt}
\bibliography{bibliography}% Produces the bibliography via BibTeX.

\appendix{
\section{Explicit form of the Carleman-Grad matrices}\label{app:I}

In this Appendix, we provide the explicit form of the matrices $A,B$, and $C$ of  Eq.~\eqref{eq:CarlemanEuler} 
for the case of a two-dimensional system. 
The Carleman vector at first order is defined as $V=(\rho,J_1,J_2,P_{11},P_{12},P_{22})$, and the 
components $V_a$ are indexed with $a$ running from $0$ to $5$. 

The final result is: 

\begin{equation}
A = dt\begin{pmatrix}
1/dt & \partial_1 & \partial_2 & 0         & 0             & 0\\
0 &            1/dt & 0            & \partial_1& \partial_2& 0\\
0 & 0            & 1/dt            & 0           & \partial_1&\partial_2\\
\omega  c_s^2 & -3  c_2^2\partial_1 & - c_s^2\partial_2 & (1/dt-\omega) & 0 & 0\\
0 & - c_2^2\partial_2 & - c_s^2\partial_1 & 0 & (1/dt-\omega) & 0\\
\omega  c_s^2 & - c_2^2\partial_1 & -3 c_s^2\partial_2 & 0 & 0 & (1/dt-\omega) \\
\end{pmatrix}.
\end{equation}

\noindent The matrices $B$ and $C$ are highly sparse, and the non-null components are

\begin{eqnarray}
B_{311}=B_{412}=B_{522}=2dt\omega,\\
C_{3110}=C_{4120}=C_{5220}=-dt\omega.
\end{eqnarray}

Usually, the derivative operators $\partial_1, \partial_2$ have to be converted to  their discretized version $D_1,D_2$, 
which are sparse matrices themselves. 
After naming the discretized version of the matrix $A$ as $\hat{A}$, for a lattice with $G$ gridpoints, this has the explicit form

\begin{equation}\label{eq:matrixA}
\hat{A} = dt\begin{pmatrix}
\mathds{1}/dt & D_1 & D_2 & 0         & 0             & 0\\
0 &            \mathds{1}_N/dt & 0            & D_1& D_2& 0\\
0 & 0            & \mathds{1}_N/dt            & 0           & D_1&D_2\\
\omega  c_s^2\mathds{1} & -3  c_2^2D_1 & - c_s^2D_2 & (1/dt-\omega)\mathds{1}_N & 0 & 0\\
0 & - c_2^2D_2 & - c_s^2D_1 & 0 & (1/dt-\omega)\mathds{1}_N & 0\\
\omega  c_s^2\mathds{1}_N & - c_2^2D_1 & -3 c_s^2D_2 & 0 & 0 & (1/dt-\omega)\mathds{1}_N \\
\end{pmatrix},
\end{equation}

\noindent where $\mathds{1}_N, D_1$ and $D_2$ have dimension $N\times N.$  
The matrices $B$ and $C$ are changed into their $N$ gridpoints version as
\begin{eqnarray}\label{eq:matrixB}
\hat{B}=\mathds{1}_N\otimes B\\\label{eq:matrixC}
\hat{C}=\mathds{1}_N\otimes C\nonumber.
\end{eqnarray}
}

\end{document}